\documentclass[conference]{IEEEtran}

\usepackage{times}
\usepackage{amsmath}
\usepackage{amssymb}
\usepackage{algorithm}
\usepackage{algorithmic}
\usepackage{graphicx}
\usepackage{subfigure}
\usepackage{verbatim}
\usepackage[usenames]{color}
\usepackage{cite}
\usepackage{url}
\usepackage{soul}
\usepackage{todonotes}

\begin{document}

\title{Mobile Anchor Assisted Node Localization for Wireless Sensor Networks }

\author{
Hongyang Chen$^{1,2}$, Qingjiang Shi$^3$, Pei Huang$^4$, H.Vincent Poor$^5$, and Kaoru Sezaki$^{1,2}$\\
$^1$Institute of Industrial Science, The University of Tokyo, Tokyo, Japan\\
$^2$CREST, JST, Japan\\
$^3$Department of Electronic Engineering, Shanghai Jiao Tong University, Shanghai, China\\
$^4$Department of Computer Science and Engineering, Michigan State University, USA\\
$^5$Department of Electrical Engineering, Princeton University, Princeton, NJ, USA}

\maketitle

\begin{abstract}
In this paper, a cooperative localization algorithm is proposed that considers the existence of obstacles in mobility-assisted wireless sensor networks (WSNs). In this scheme, a mobile anchor (MA) node cooperates with static sensor nodes and moves actively to refine location performance. The localization accuracy of the proposed algorithm can be improved  further by changing the transmission range of mobile anchor node. The algorithm takes advantage of cooperation between MAs and static sensors while, at the same time, taking into account the relay node availability to make the best use of beacon signals. For achieving high localization accuracy and coverage, a novel convex position estimation algorithm is proposed, which can effectively solve the localization problem when infeasible points occur because of the effects of radio irregularity and obstacles. This method is the only range-free based convex method to solve the localization problem when the feasible set of localization inequalities is empty. Simulation results demonstrate the effectiveness of this algorithm.
\end{abstract}


%
\IEEEpeerreviewmaketitle

\section{Introduction}

\PARstart{L}{ocalization} algorithms for wireless sensor networks (WSNs) have been designed to find sensor location information, which is a key requirement in many applications of WSNs. Generally speaking, based on the type of information required for localizaiton, protocols can be divided into two categories: (i) range-based and (ii) range-free protocols [1]. Range-based techniques require special hardware for estimating the distance between anchors and sensors, which may become prohibitively expensive for large networks [2]. Range-free techniques, on the other hand, do not impose such complexity as the anchor informs other sensors about its own position through message passing [3]. After finishing the distance-from-anchor estimation process, a regular sensor can determine its own position, through a variety of methods, such as multilateration, triangulation, etc. If necessary, an optional step is performed, in which regular sensors exchange messages among themselves to refine their locations. Due to the hardware limitations and power constraints of sensors, solutions of range-free localization are often preferable and can be considered as cost-effective options when compared with more expensive and energy-consuming range-based schemes [4]-[7]. In this paper, we focus on the investigation of range-free localization algorithms for mobility-assisted WSNs.

An obstacle can be dynamically formed due to unbalanced deployment, failure or power exhaustion of sensor nodes, animus interference, or physical obstacles such as mountains or buildings. In this paper, we consider only physical obstacles. Most previous algorithms cannot work well in anisotropic networks, where obstacles appear among sensor nodes. However, anisotropic networks with obstacles are more likely to exist in practice for several reasons. Firstly, a uniform node distribution cannot always be achieved because of random deployment of sensor nodes, which may cause some regions to not contain any sensor node. Secondly, unbalanced power consumption among nodes results in some regions without functionality of sensing and communication. Thirdly, physical obstacles such as mountains or buildings will naturally exist in many networks.

In this work, we propose a multi-power level mobile anchor assisted range-free algorithm for WSNs with obstacles. By using a relay node, our scheme can effectively reduce the effects of obstacles on node localization. Furthermore, our scheme can calculate the positions of infeasible points caused by a complex radio transmission environment, which is recognized as a problem when the feasible set for localization inequalities is empty.



\section{Collaborative Localization using A Mobile Anchor}
In this section, we propose a collaborative node localization approach using an MA. We first introduce the technical preliminaries of our algorithm in subsection A and then formulate the localization problem as an optimization problem in subsection B. We propose an algorithm for decreasing the impact of obstacles in subsection C.

\subsection{Background }
In WSNs, a node can determine whether it is in the transmission radius of an anchor node according to the beacon signal received from the one-hop anchor. The anchor node can adjust its transmission radius by tuning the transmission power [9]. For example, the TelosB mote is equipped with an IEEE 802.15.4 compliant Chipcon CC2420 radio, which has 31 transmission power levels between -25 and 0 dBm.

We have conducted experiments on a testbed composed of TelosB motes. As shown in Fig.1, the experiments demonstrate that the transmission radius of a sensor can be efficiently changed by tuning the transmission power level.

\begin{figure}[t]
\centering
\includegraphics[width=3.5in]{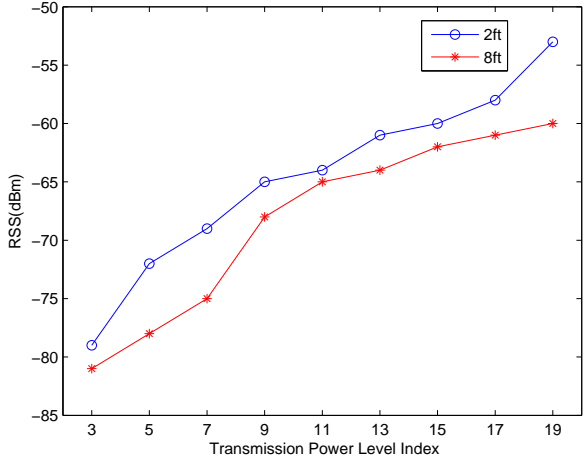}
\caption{Experimental RSSI measurements} \label{fig:RSSI measurements}
\end{figure}

We assume an anchor node has $M$ levels of transmission power, and the related transmission range is $R_i$, $i=1,2, \ldots ,M$. Normally, the MA is assumed to have a global positioning system (GPS) receiver and knows its position. During the moving period, the MA transmits beacon signals at varying power levels consecutively including its ID, current position, transmission power and transmission radius. After receiving these beacon signals, an unknown-position sensor can construct an effective constraint on its position.

For example, we assume that the current position for the MA is $a$ and its transmission radius is $R$. If the unknown-position sensor, at position $x$, receives the beacon signal, we can conclude that the distance between both nodes satisfies
\begin{eqnarray}
|| {x - a} || \le R.
\end{eqnarray}
Otherwise,
\begin{eqnarray}
|| {x - a} || > R.
\end{eqnarray}
Using the various transmission power levels of the MA in different positions, the unknown-position sensor can obtain a set of inequalities on $x$:
\begin{eqnarray}\label{eq:quadratic inequalities}
r_i  < || {x - a_i }|| \le R_i, ~~~i=1,2,...,M
\end{eqnarray}
where $a_i$ is the position of the MA at time $i$, $r_i$ (it might be zero) and $R_i$ are \emph{valid radii} for that time. Herein, the valid constraint radii denote the related lower and upper bounds, for the tightest constraint among all of the constraints that are constructed by all of the transmission powers for the mobile anchor node at position $a_i$.


Hence, the localization problem based on an MA with variable transmission power can be successfully converted into the problem of solving a set of quadratic inequalities (\ref{eq:quadratic inequalities}). Some of the location algorithms (eg, [6] and [16]) are also based on the solution of a set of quadratic inequalities. However, their methods all assume that the set of quadratic inequalities (\ref{eq:quadratic inequalities}) must have solutions. Nevertheless, because of the complicated transmission environment, there are two different location scenarios as shown in Fig. 2: the set of quadratic inequalities has a solution (i.e., the feasible set is nonempty) for the first case; the set of quadratic inequalities have no solution (i.e., the feasible set is empty) for the second case. The dots on the figure represent the anchors and the squares represent the unknown-position sensor.

\begin{figure}[t]
\centering
\includegraphics[width=2.5in]{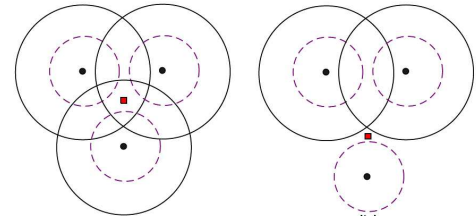}
\caption{Two localization scenarios: (a) The feasible set is nonempty; (b) The feasible set is empty} \label{fig:RSSI measurements}
\end{figure}

For the above two different scenarios, we propose a novel
localization algorithm based on convex optimization to solve the
problem when the feasible set is empty. To the best of our
knowledge, our proposed method is the only range-free algorithm
using convex optimization to solve the problem when the feasible set
is empty.

\subsection{Localization Algorithm Using Convex Optimization}
In real environments, the actual transmission radius varies in
different directions of radio propagation because of the
non-isotropic properties of the propagation medium and the
heterogeneous properties of devices. It is possible that there is no
communication between two nodes although their relative distance is
within their ideal transmission radius. On the other hand, two nodes
may also be able to communicate although their relative distance is
larger than their transmission radius. Thus, with the effects of
radio irregularity and obstacles, a localization algorithm might not
be able to guarantee full coverage and an infeasible case would
occur [10]-[14].

In order to deal with the case with an empty feasible set, we propose a
novel convex position estimation algorithm, which can provide good
position estimation accuracy in both the feasible case and the
infeasible case.

As shown in Fig.~3, for the single constraint case ($r < || {x -
a}|| < R$), it is easy to see that the optimal position estimate
lies on the circle with center $a$ and radius $\frac{{R + r}}{2}$.
In the figure, the square indicates the possible position for the
optimal position estimate and the black dot denotes the anchor node
with position $a$.

\begin{figure}[t]
\centering
\includegraphics[width=2.5in]{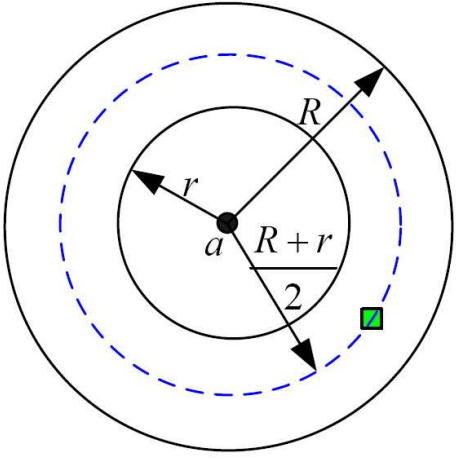}
\caption{The single constraint case} \label{fig:single constraint}
\end{figure}

The position estimate can be found by minimizing the following expression:
\[(|| {x - a}|| - r)^2  + (||x - a|| - R)^2.\]
Inspired by the single constraint case, for the inequalities under
multiple constraints, we can probably find the optimal position
estimate by solving the following problem:
\begin{eqnarray}\label{eq:multiple constraints equation}
\mathop {\min }\limits_x \sum\limits_i {\left[ {(||x - a_i || - r_i )^2  + (||x - a_i || - R_i )^2 } \right]}.
\end{eqnarray}
%
%
Obviously, the problem (\ref{eq:multiple constraints equation}) is
nonconvex. Moreover, this problem cannot be directly approximated by
using some convex relaxation techniques like that of [15].
To approximately solve the problem via convex relaxation techniques,
we transform it to the following problem\footnote{The possible case
where there is only the lower bound $r_i$ or the upper bound $R_i$
(e.g, Fig. 2(b)) is not considered in this formulation. However,
such a case can still be handled by using the same convex relaxation
technique.}:
\begin{eqnarray}\label{eq:convex transform}
\mathop {\min }\limits_x \sqrt {\sum\limits_i {\left[ {(||x - a_i ||^2  - r_i ^2 )^2  + (||x - a_i ||^2  - R_i ^2 )^2 } \right]} }~.
\end{eqnarray}
Although, the problem (\ref{eq:convex transform}) is still
nonconvex, we can turn it into a convex problem by using a convex
relaxation technique. Firstly, the problem (\ref{eq:convex
transform}) can be transformed to the epigraph [8],
\begin{eqnarray}\label{eq:epigraph}
\mathop {\min }\limits_{x,v,t}~~ t \nonumber \\
{\text{s}}{\text{.t}}{\text{.}}~||v|| \le t \nonumber \\
||x-a_i||^2 - r_i^2=v_{i1} \nonumber \\
||x-a_i||^2 - R_i^2=v_{i2}, ~~~~\forall i
\end{eqnarray}
where $v = \left[ {\begin{array}{*{20}c}
{v_{11} } & {v_{12} } &  \cdots  & {v_{i1} } & {v_{i2} } &  \cdots  & {v_{n1} } & {v_{n2} }  \\
\end{array}} \right]^T$. It is easy to see that,
\begin{eqnarray}\label{eq:easily find}
\begin{array}{l}
 ||x - a_i ||^2  = \left[ {1 - a_i ^T } \right] \times %
 \left[ {\begin{array}{*{20}c}{x^T x} & {x^T }  \\
 x & {I_2 }  \\ \end{array}} \right] \times %
 \left[ {\begin{array}{*{20}c} 1  \\ { - a_i }  \\ \end{array}} \right] \\
\hspace{ 16mm }= A_i  \bullet Y \\
\end{array}
\end{eqnarray}
where $A_i  = \left[ {\begin{array}{*{20}c}
   1  \\
   { - a_i }  \\
\end{array}} \right]\left[ {1 - a_i ^T } \right]$,~%
$Y = \left[ {\begin{array}{*{20}c}
   y & {x^T }  \\
   x & {I_2 }  \\
\end{array}} \right]$,~%
$y=||x||^2$ and ~%
$I_2  = \left[ {\begin{array}{*{20}c}
   1 & 0  \\
   0 & 1  \\
\end{array}} \right]$. Here ``$\bullet$''denotes the inner product, namely, the sum of the products of corresponding elements of two matrices.

Using (\ref{eq:easily find}), the equivalent form for the problem
(\ref{eq:epigraph}) is obtained as follows:
\begin{eqnarray}\label{eq:equivalent epigraph}
 \mathop {\min }\limits_{Y,v,t} ~~ t \nonumber\\
{\text{s}}{\text{.t}}{\text{.}} ~||v|| \le t \nonumber\\
 Y_{2:d + 1,2:d + 1}  = I_2 \nonumber \\
 A_i  \bullet Y - v_{i1}  = r_i ^2  \nonumber\\
 A_i  \bullet Y - v_{i2}  = R_i ^2,~~~   \forall i \nonumber\\
 y = ||x||^2~.
\end{eqnarray}
In (\ref{eq:equivalent epigraph}), $||v|| \le t$ is a second-order
cone. However, the problem (\ref{eq:equivalent epigraph}) is still
not convex due to the nonlinear equality constraint. Herein, we
relax the equality $y = ||x||^2$ to $y \ge ||x||^2$ which is
equivalent to requiring that \emph{Y} is a positive semidefinite matrix by the Schur
complement theorem [8]. Hence, we have the following convex
optimization problem:
\begin{eqnarray}\label{eq:convex optimization}
 \mathop {\min }\limits_{Y,v,t} ~~ t \nonumber\\
{\text{s}}{\text{.t}}{\text{.}} ~||v|| \le t \nonumber\\
 Y \ge 0 \nonumber\\
 Y_{2:d + 1,2:d + 1}  = I_2 \nonumber \\
 A_i  \bullet Y - v_{i1}  = r_i ^2  \nonumber\\
 A_i  \bullet Y - v_{i2}  = R_i ^2, ~~~  \forall i
\end{eqnarray}
where $Y \ge 0$ indicates that $Y$ is a positive semidefinite
matrix. The resulting problem is a convex cone programming problem
which can be solved by using efficient interior-point algorithms
[8]. After obtaining the value of $Y$, we can further calculate the
position estimate for the unknown-position sensor $x$, namely, $x =
Y_{2:3,1:1}$, where $Y_{2:3,1:1}$ denotes the vector constructed
from the elements of the second and third rows for the first column
of the matrix $Y$.

\subsection{Algorithm for Decreasing the Impact of Obstacles}
In this paper, we assume that boundary nodes around the obstacle have been discovered by some boundary recognition algorithms [17], so that each sensor node knows whether it is a boundary node or not. Only boundary nodes can participate in contention for relaying beacons from the MA because their rebroadcasts may cover some blind areas as shown in Fig. 4. Hearing a beacon from the MA, boundary nodes will compete to relay this location information through a distributed contention process. The probability that a candidate node wins the contention depends on the node's remaining energy and the number of neighboring sensors. A node with greater remaining energy and greater number of neighbors has higher priority to be the optimal relay node. The proposed selection scheme for the optimal relay node is concluded as follows:

Receiving a beacon from the MA, a boundary node sets a backoff timer that defines the amount of time that the node must wait before rebroadcasting the location information. The backoff time $\delta$ is calculated as
\begin{eqnarray}\label{eq: backoff time}
\delta  = (\alpha (used\_energy/initial\_energy) + \nonumber\\
\beta /num\_neighbors) * max\_delay
\end{eqnarray}
where $\alpha$ and $\beta$ are coefficients that provide different weights for different parameters.
The specific values of $\alpha$ and $\beta$ can be set depending on which property is more important for users: energy balance or coverage efficiency. In total, $\alpha  + \beta  = 1$.
We can see that a greater remaining energy and a greater number of neighbors will lead to a shorter backoff time. If a candidate boundary node does not hear any beacon signal from other sensors during its backoff time, it will rebroadcast the beacon signal and other boundary nodes will cancel their contentions if they receive the rebroadcast of the beacon. As a result, the node with the highest priority will rebroadcast first and win the competition to serve as the relay for the MA's beacon signal. In this way, we can deliver the MA's location information to some areas that cannot receive the MA's direct communication. Similarly to (3), the unknown-position sensor in these special areas can obtain a set of inequality constraints on $x$ :
\begin{eqnarray}\label{eq:quadratic inequalities 2}
r_i  < || {x - a_i }|| \le R_i+R_{relay}, ~i=1,2,...,n
\end{eqnarray}
where $R_{relay}$ is the current transmission radius for the relay node.
We can also use the proposed convex localization algorithm to solve the problem (\ref{eq:quadratic inequalities 2}).  Based on this scheme, we can efficiently decrease the impact of the obstacle on node localization and improve the location accuracy.

\begin{figure}[t]
\centering
\includegraphics[width=2.5in]{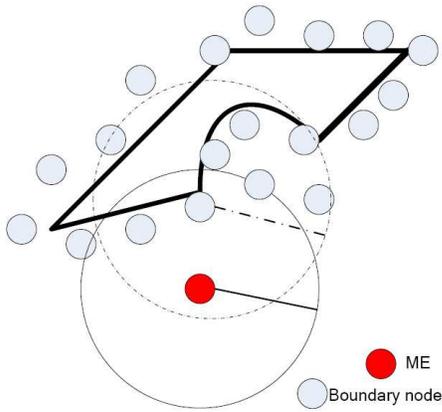}
\caption{A sensor network with an obstacle} \label{fig:Sensor networks with obstacle}
\end{figure}

\section{   NUMERICAL RESULTS}
In this section, simulation results are presented and analyzed. The performance evaluation focuses on the position estimation accuracy of the proposed algorithm. We consider a 2-dimensional region with a size of 100 m x 100 m. We assume the MA has two level transmission power with the transmission radii $r$ and $R=2r$, respectively. The method of [16] is also evaluated with our proposed algorithm for performance comparisons. First, we deploy 100 sensor nodes randomly and the transmission radius  $r$ is set to 15 meters. In subsection \emph{A}, we simulate our algorithm in the ideal situation. The effects of degree of irregularity (DOI) and obstacle on localization performance will be discussed in subsection \emph{B}. All simulation results are averaged over 100 network scenarios.
The average localization error is used to evaluate the performance for our localization algorithm. Localization error is defined as follows:
 \begin{eqnarray}
error = \frac{1}{N}\sum\limits_{i = 1}^N {\left\|{{x_i} - {{\hat x}_i}}\right\|}\times \frac{1}{r},
 \end{eqnarray}
where $x_i$ is the actual position for node $i$  and ${\hat x}_i$ is the estimated position of node $i$. Note that we normalize the absolute localization error using radio range.
For instance, an error of 20\% means that the localization error is 20\% of the radio range.

\subsection{Performance in the Ideal Environment}
In this subsection, we give the simulation results for different algorithms in the ideal situation, namely, when there is no obstacle in the sensing area. We use the DOI to indicate the radio irregularity characteristic.
Fig. 5(a) and 5(b) shows the simulation results in the ideal situation, where the true nodes are denoted by circles, the position estimates are denoted by asterisks, and the lines that link the true nodes and the estimates represent the estimation errors. It is clear from Fig. 5(a) and 5(b) that our algorithm works better than the algorithm of [16] in terms of the average localization error.



\begin{figure}[t]
\centering \subfigure[ ]{
\label{fig:subfig:our method} 
\includegraphics[width=0.43\textwidth]{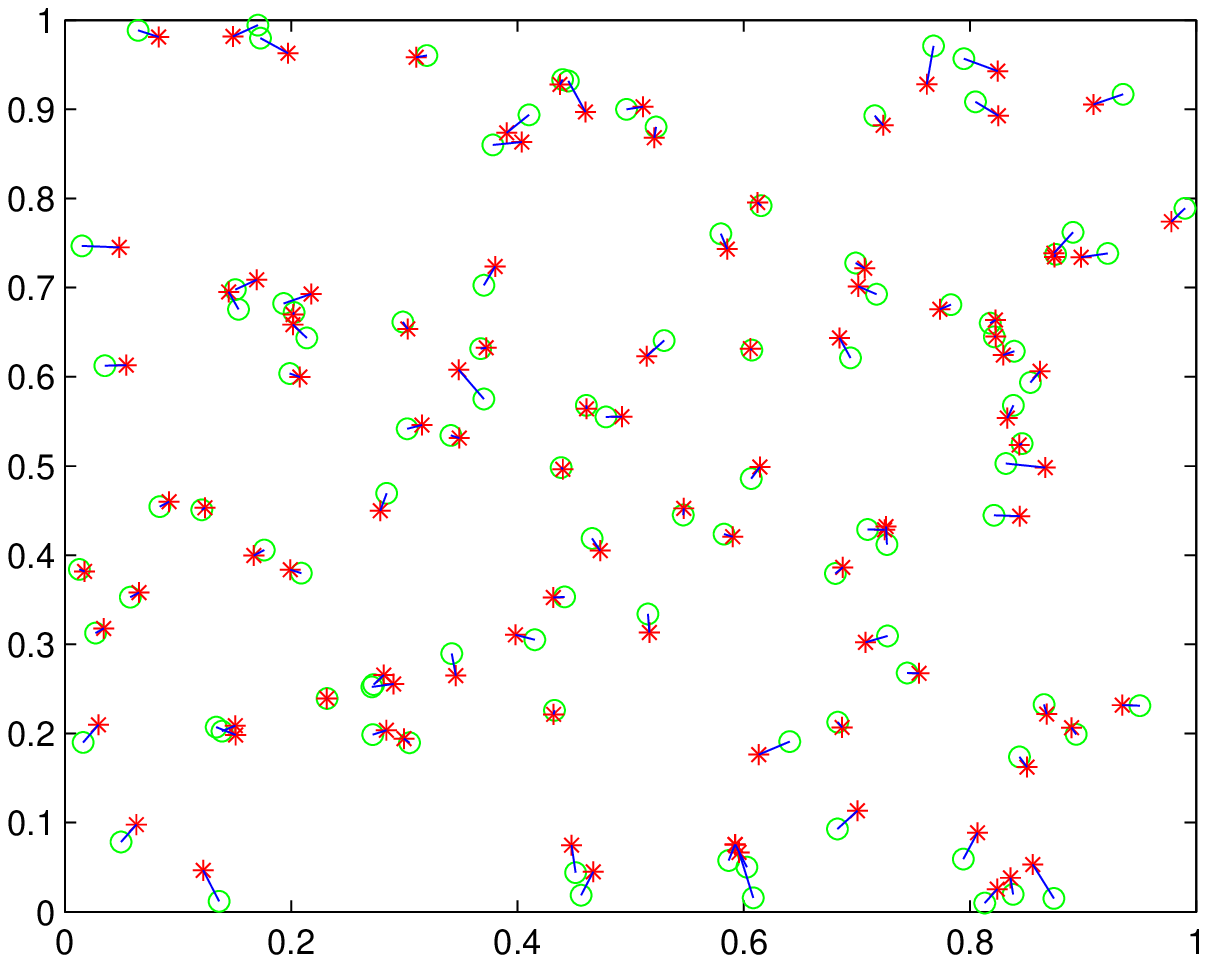}}
\hspace{0in} \subfigure[]{
\label{fig:subfig:method [16]} 
\includegraphics[width=0.43\textwidth]{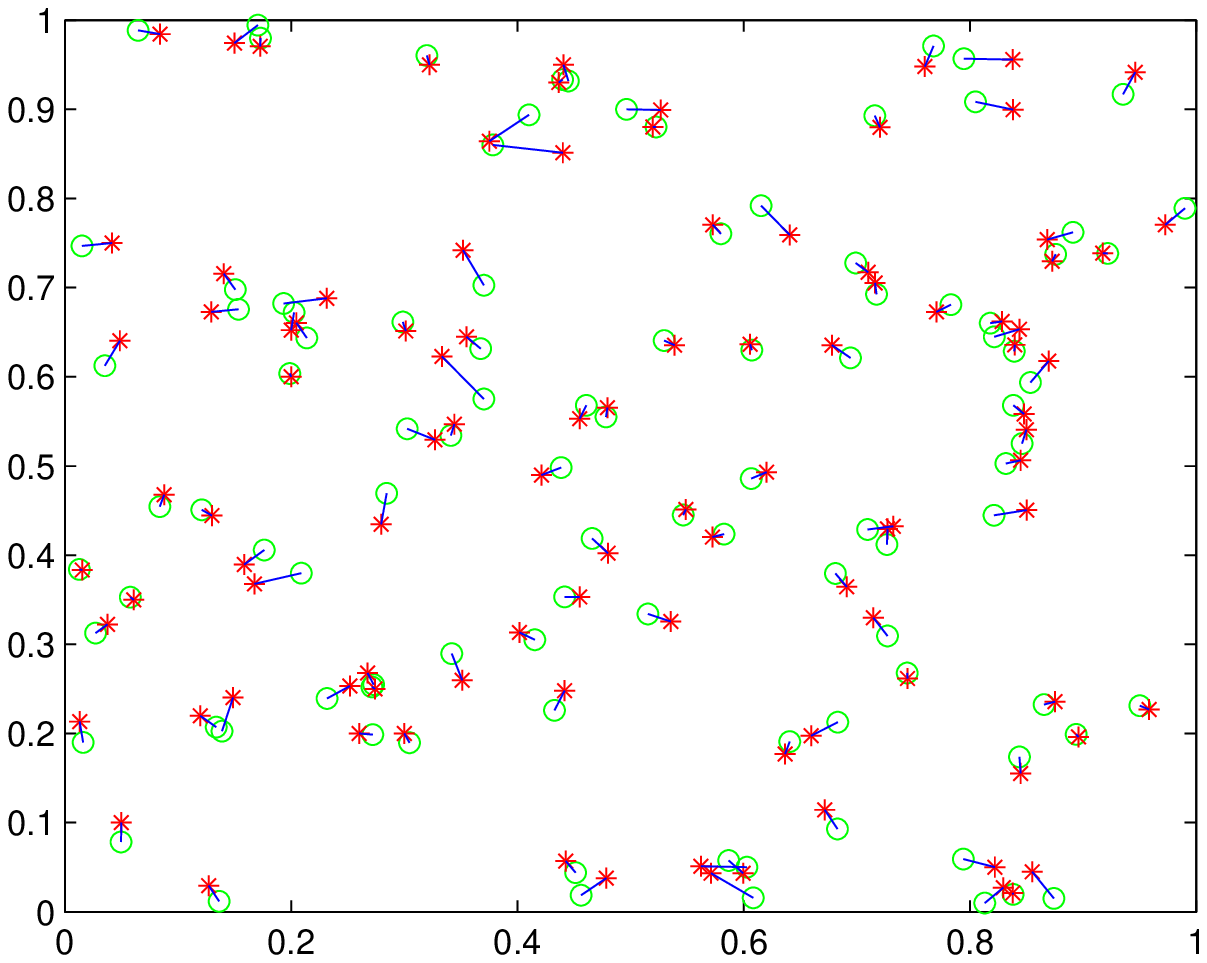}}
\caption{Performance comparison: (a) Localization error of our method (DOI=0.2, $\textrm{error}=11.68\%$); (b) Localization error of method [16] (DOI=0.2, $\textrm{error}=13.7\%$)}
\label{fig:location error} 
\end{figure}

\subsection{Performance in the Non-ideal Environment }
For the next set of experiments, we use a fading coefficient \emph{(f)} that represents the percentage of total mobile beacon points that cannot be heard by the sensor at any given time. This models the obstacles encountered in the sensing area that limit the number of mobile beacon points that can be heard at any point. As Fig. 6(a) and 6(b) illustrates, our algorithm outperforms the algorithm of [16] in terms of the average localization error in this non-ideal environment.

\begin{figure}[t]
\centering \subfigure[ ]{
\label{fig:subfig:proposed movement} 
\includegraphics[width=0.43\textwidth]{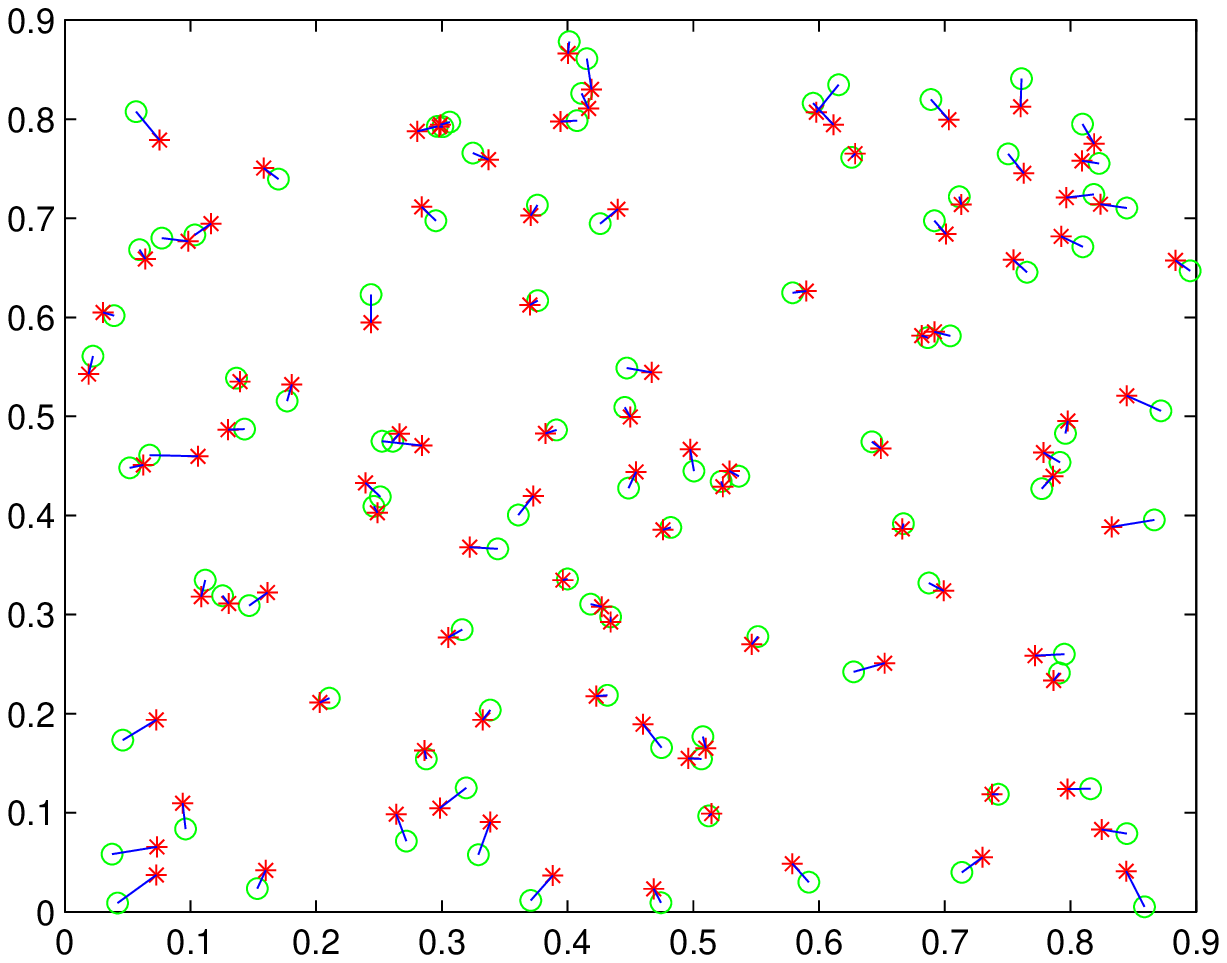}}
\hspace{0in} \subfigure[]{
\label{fig:subfig:random movement} 
\includegraphics[width=0.43\textwidth]{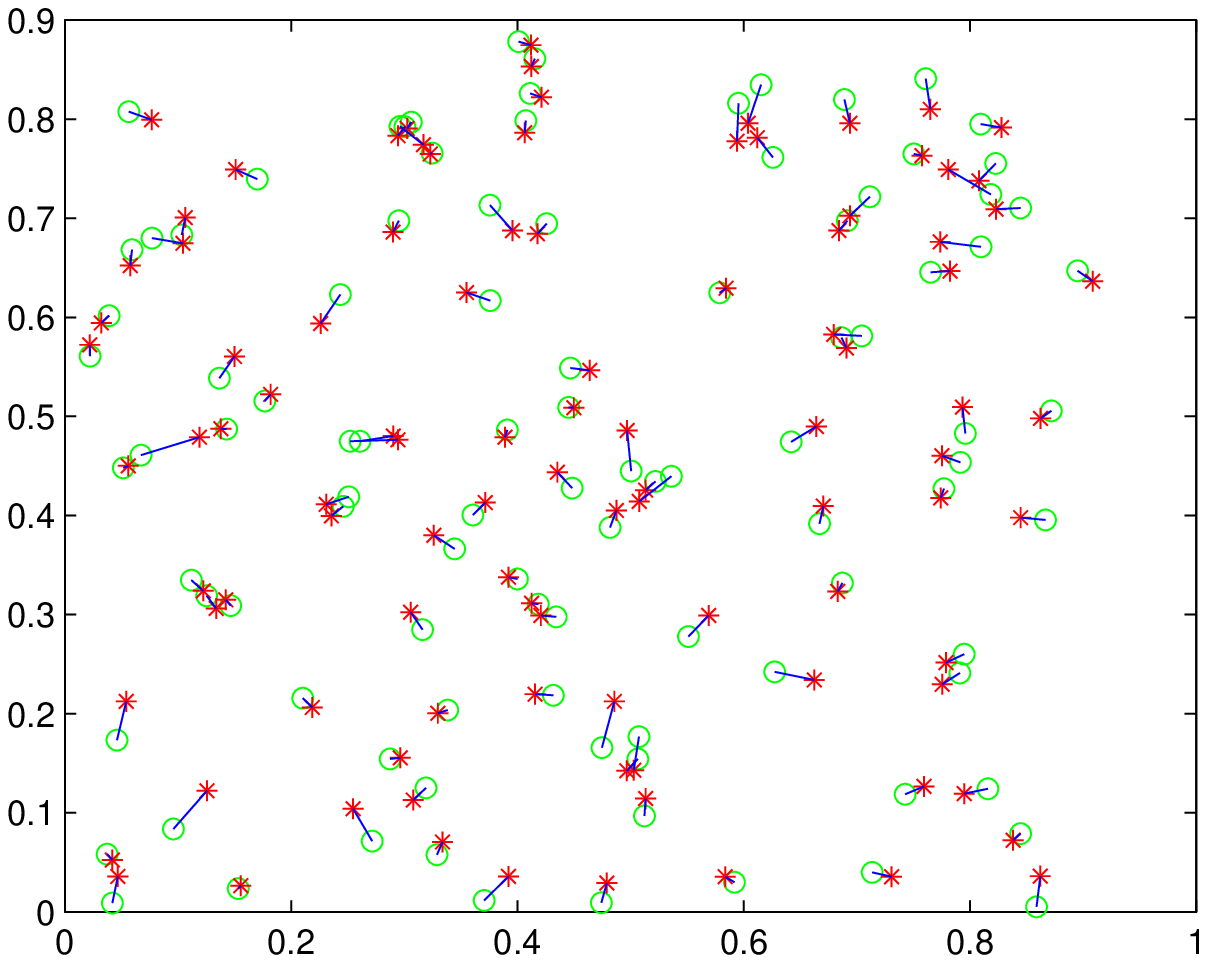}}
\caption{Performance comparison in the non-ideal environment: (a) Localization error of our method (DOI=0.2, \emph{f}=0.1, $\textrm{error}=12.95\%$); (b) Localization error of method [16] (DOI=0.2, \emph{f}=0.1, $\textrm{error}=14.89\%$)}
\label{fig:location error} 
\end{figure}

\section{Conclusions}
We have presented a new cooperative localization scheme that can achieve high localization accuracy in mobility-assisted wireless sensor networks when obstacles exist. Considering the complex localization scenario, namely, the feasible set is empty, a convex localization algorithm has been presented to address the effects of non-ideal transmission of radio signals. It has been shown in the simulation results that the proposed cooperative localization scheme can significantly improve the localization accuracy by including a mobile element. In future work, we intend to verify and improve the proposed cooperative localization scheme using real sensors in a mobility-assisted wireless sensor networks.

\section*{Acknowledgement}
This work was supported by the CREST Advanced Integrated
Sensing Technology project of the Japan Science and Technology
Agency.

\end{document}